\documentclass[onecolumn,traditabstract]{aa}  
\usepackage{graphicx}
\usepackage{txfonts}
\usepackage{color}
\usepackage[]{xcolor}
\usepackage{overpic}
\usepackage{tabularx,colortbl}
\newcommand{\ca}{\mbox{Ca\,{\sc ii}~K\,}}
\definecolor{darkergrey}{rgb}{0.65,0.65,0.65}

\begin{document}
\authorrunning{Chatzistergos et al.}
\titlerunning{Long-term changes in solar activity and irradiance}
\title{Long-term changes in solar activity and irradiance}
\author{Theodosios~Chatzistergos, Natalie~A.~Krivova, Kok~Leng~Yeo}
\offprints{Theodosios Chatzistergos  \email{chatzistergos@mps.mpg.de}}
\institute{Max Planck Institute for Solar System Research, Justus-von-Liebig-weg 3, 37077 G\"{o}ttingen, Germany }

\abstract{The Sun is the main energy source to Earth, and understanding its variability is of direct relevance to climate studies.
Measurements of total solar irradiance (TSI) exist since 1978, but this is too short compared to climate-relevant time scales.
Coming from a number of different instruments, these measurements require a cross-calibration,
which is not straightforward, and thus several composite records have been created.
All of them suggest a marginally decreasing secular trend in solar minima levels since 1996. Most composites actually feature a weak decrease over the entire period of observations, which is also seen in observations of the solar surface magnetic field and is further supported by \ca data. Some inconsistencies, however, remain and overall the magnitude and even the presence of the secular change over the satellite era remain uncertain.
Different models have been developed, which are used to understand the irradiance variability over the satellite period and to extend the records of solar irradiance back in time.
Differing in their methodologies, all models require proxies of solar magnetic activity as input.
The most widely used proxies are sunspot records and cosmogenic isotope data on centennial and millennial time scale, respectively.
None of this, however, offers a sufficiently good, independent description of the long-term evolution of faculae and network responsible for solar brightening.
This leads to significant uncertainties in the amplitude of the changes in solar irradiance  on time scales longer than the solar cycle.
Here we review recent efforts and advances aiming at improving long-term irradiance reconstructions and to reduce the existing uncertainty in the magnitude of the long-term variability. 
In particular, by employing state-of-the-art 3D magnetohydrodynamical simulations, an upper limit of $2\pm0.7$~W\,m$^{-2}$ was set on the possible increase of TSI since the end of the Maunder minimum as compared to the 2019 minimum level. Besides, significant progress has been made in collecting and processing historical solar observation in \ca spectral line, which provide direct information on bright magnetic features on the Sun and can be used to improve the accuracy of past irradiance reconstructions, pending some remaining issues with the data. 
}

\maketitle

\section{Introduction}
\label{sec:intro}
\sloppy

In view of the recent rapid global warming of Earth \citep{masson-delmotte_annex_2021}, understanding mechanisms driving the climate variability is a challenging while pressing endeavour.
Being the dominant supplier of energy to Earth 
 \citep[see, e.g.,][]{kren_where_2017}, the Sun is also one of the main actors.
Various mechanisms of solar influence on Earth's climate have been proposed, which involve variability in the total and spectral solar irradiance, as well as fluctuations in the fluxes of the galactic cosmic rays (GCRs) and solar energetic particles (SEPs) \citep[see, e.g.,][for reviews]{haigh_sun_2007,gray_solar_2010,lockwood_solar_2012}.

The total solar irradiance (TSI) is the spectrally integrated radiative energy flux incident on the top of the Earth's atmosphere at the mean Sun--Earth distance of 1~a.u., and it describes the total radiative energy of the Sun received by Earth's system.
TSI has been regularly monitored from space since the late 1970s and was found to
vary on all time scales covered by direct measurements \citep[e.g.][]{hickey_initial_1980,willson_solar_1988,frohlich_solar_2006,frohlich_total_2012,kopp_impact_2016}.
It is also expected to vary on longer time scales.
For example, from sunspot observations and cosmogenic isotope data we know that the Sun was in a state of enhanced activity (called Modern grand maximum) over the second half of the 20th century \citep{solanki_unusual_2004,chatzistergos_new_2017,usoskin_history_2023}, while solar activity was somewhat lower during the period called Dalton minimum \citep[1790–-1820;][]{usoskin_history_2023,hayakawa_stephan_2021} and considerably lower during the Maunder minimum \citep[1645–-1715;][]{eddy_maunder_1976,usoskin_maunder_2015}.
The Maunder minimum and the Modern grand maximum are the only so-called grand extrema of solar activity, which are covered by direct solar observations, while concentrations of cosmogenic isotopes evidence the existence of other such events in the past \citep{usoskin_solar_2021,usoskin_history_2023}. 
On time scales relevant to climate studies, the driver of the irradiance variability is the solar surface magnetic field \citep{krivova_reconstruction_2003,domingo_solar_2009,shapiro_nature_2017,yeo17b}.

The variability strongly depends on the wavelength, growing considerably towards the UV part of the spectrum \citep{rottman_observations_1988,floyd_11_2003}. 
The solar radiative energy flux per unit wavelength or within a given spectral interval is called spectral solar irradiance (SSI).
SSI has also been monitored over roughly the same period, although with significant gaps in the wavelength and temporal coverage \citep[][and references therein]{ermolli_recent_2013,deland_creation_2019,woods_overview_2021}. 
While the longwave (mainly visible and  infrared, but also parts of UV above 320 nm) radiation penetrates down to the lower layers of the atmosphere and the surface, heating these directly,
the UV irradiance heats the upper and middle terrestrial atmosphere and plays a critical role in chemical processes there.
The signal penetrates down to the lower layers through dynamical coupling mechanisms 
\citep{haigh_role_1994, haigh_sun_2007,gray_solar_2010}.

Not only solar radiation can influence the terrestrial atmosphere and climate.
Also SEPs affect the ionisation state and thus the chemical processes in the upper and middle atmosphere \citep[see, e.g., the reviews by ][and references therein]{sinnhuber_energetic_2012,sinnhuber_energetic_2020,mironova_energetic_2015}. 
Furthermore, yet more energetic GCR particles can penetrate down to the lower layers of the atmosphere (lower  stratosphere and the troposphere).
Although the link between the GCR-caused ionisation of the lower atmosphere and cloud formation has been speculated it has not been confirmed \citep{kulmala_atmospheric_2010,laken_cosmic_2009,pierce_can_2009,laken_cosmic_2012,calogovic_sudden_2010}, while results from the Cosmics Leaving OUtdoor Droplets (CLOUD) experiment in CERN \citep{dunne_global_2016,pierce_cosmic_2017} suggest that GCRs do not have any appreciable effect on cloud formation.
The GCR flux at Earth is modulated by the solar magnetic field, being weaker at higher activity, in contrast to the flux of energetic particles originating from the Sun itself.
 
Whichever of these mechanisms or their combination are in play, the fundamental driver of all these changes is the solar magnetic field.
Thus understanding the long-term changes of the solar surface magnetic field and solar magnetic activity is key to understanding the solar influence on Earth's climate. This review provides a short update on the recent efforts to reconstruct past solar magnetic activity.
We focus on the recovery and reconstructions of long-term solar activity proxies, such as sunspot and \ca plage observations, as well as constraining the long-term changes in solar irradiance. 

In Sect.~\ref{sec:satellite} we discuss measurements of solar irradiance and solar surface magnetic field over the satellite era.
We give an overview of the various irradiance reconstructions and proxies of past solar magnetic activity used by these reconstructions in Sect. \ref{se:recs}, while in Sect. \ref{sec:dimmeststate} we briefly describe the recent model constraining the possible range of solar irradiance variations.
Finally, in Sect. \ref{sec:summary} we summarise and draw our conclusions.

\section{Satellite era}
\label{sec:satellite}

\subsection{TSI measurements and their composites}
\label{sec:meas}

While the links between solar activity and Earth's climate have been speculated for a long time \citep[][cf. \citealt{love_insignificance_2013,chatzistergos_is_2023}]{herschel_observations_1801,eddy_maunder_1976}, quantitative studies have basically started with the realisation that the ``solar constant'', as TSI was called at earlier times, was not constant.
This only became possible with the start of space-borne monitoring of solar irradiance in 1978 \citep{hickey_initial_1980,willson_solar_1981,willson_variations_1981}. These measurements revealed variability on all observable time scales, most prominent being the short-term (time scales of days to weeks) fluctuations with amplitudes of up to 0.3\% and the variability in phase with the solar activity cycle by roughly 0.1\% during the last decades covered by satellite data \citep{willson_suns_1991,frohlich_evidence_2009,frohlich_total_2012,kopp_impact_2016}.
Being nearly continuous since 1978, space-borne TSI measurements were not carried out by a single instrument, but by
multiple partially overlapping experiments \citep[see, e.g.,][]{kopp_assessment_2014,kopp_magnitudes_2016,montillet_data_2022}. 
Due to different absolute scales \citep{kopp_total_2012}, various instrument-specific degradations \citep{frohlich_-flight_1997,frohlich_observations_2000,frohlich_solar_2006,dewitte_measurement_2004,kopp_science_2021}, or continuity issues \citep[e.g.,][see also the review by \citealt{zacharias_independent_2014}]{lee_iii_long-term_1995,krivova_acrim-gap_2009}, cross-calibration of the individual data and construction of a continuous composite is non-trivial.
A lot of effort went into this task, which are extensively discussed elsewhere \citep[e.g.,][]{willson_composite_2003,dewitte_measurement_2004,frohlich_solar_2006,kopp_assessment_2014,zacharias_independent_2014,dudok_de_wit_methodology_2017,montillet_data_2022}.
Here, we will only briefly summarise the current status.

Three ``classical'' and most widely cited composites have been developed by the respective instrumental teams, these are:
ACRIM\footnote{Available at \url{https://web.archive.org/web/20170611210135/http://acrim.com/}} \citep[Active Cavity Radiometer Irradiance Monitor, which is the instrument taken as the reference by][]{willson_total_1997,willson_composite_2003}, PMOD\footnote{Available at \url{https://www.pmodwrc.ch}} \citep[named after Physikalisch-Meteorologisches Observatorium Davos;][]{frohlich_solar_2006}, and ROB\footnote{Available at \url{https://www.sidc.be/observations/space-based-timelines/tsi}} \citep[named after Royal Observatory of Belgium, previously referred to as RMIB, Royal Meteorological Institute of Belgium, or IRMB in French;][]{dewitte_total_2004,dewitte_total_2016,dewitte_centennial_2022}.
Recently, \citet{schmutz_changes_2021} revised and updated the PMOD composite.
In particular, he
averaged the PMOD series with SORCE/TIM data over 2003--2017 and extended it with SORCE/TIM and TSIS/TIM data afterwards.
PMOD and the series by \citet{schmutz_changes_2021} are identical prior to 2003, however the latter has only monthly values.
Also, \citet{scafetta_empirical_2023} have slightly revised the ACRIM composite, extending it until 2022 with the average values from the SOHO/VIRGO$^3$ \citep[Variability of solar IRadiance and Gravity Oscillations experiment on-board the SOlar and Heliospheric Observatory;][]{frohlich_-flight_1997}, SORCE/TIM \citep[Total Irradiance Monitor on-board the Solar Radiation and Climate Experiment mission;][]{kopp_science_2021}, and TSIS-1/TIM \citep[Total and Spectral Solar Irradiance Sensor][]{pilewskie_tsis-1_2018} TSI data.

\begin{figure*}[]
	\centering
	\includegraphics[width=0.95\linewidth]{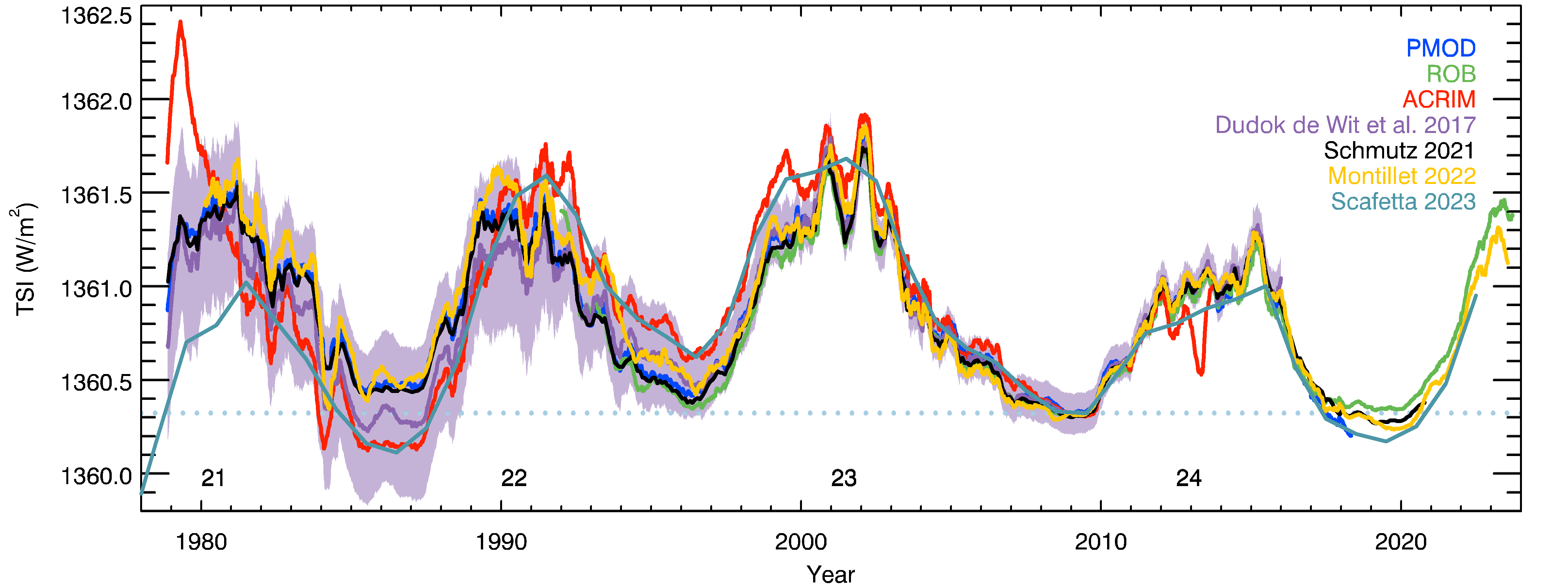}
	\caption{Composites of TSI measurements. Shown are the PMOD (blue), ROB (green), ACRIM (red), \citet[][purple]{dudok_de_wit_methodology_2017}, \citet[][black]{schmutz_changes_2021}, \citet[][yellow]{montillet_data_2022}, 
	and \citet[][pale blue]{scafetta_empirical_2023} series.  
	Purple shaded area marks the 1$\sigma$ uncertainty level for the \citet[][]{dudok_de_wit_methodology_2017} TSI composite. All series are offset to match the PMOD composite during the 2009 activity minimum. Shown are 180-day running means, with the exception of the composites by \citet[][]{scafetta_empirical_2023}, which is provided as annual mean values, and 
	by \citet[][]{schmutz_changes_2021}, provided as monthly means and shown here as 6-month running means. The light blue dotted horizontal line marks the PMOD TSI value over 2009. The numbers in the lower part of the figure denote the conventional solar cycle numbering.	}
	\label{fig:tsicomposites}
\end{figure*}

More recently two new series have been presented by \citet{dudok_de_wit_methodology_2017}\footnote{Update is available at \url{https://spot.colorado.edu/~koppg/TSI/}} and \citet[][Composite PMOD-Data Fusion, CPMDF, hereafter]{montillet_data_2022}\footnote{Available at \url{ftp://ftp.pmodwrc.ch/pub/data/irradiance/virgo/TSI/TSIcomposite/}}.
In contrast to the ``classical'' composites, which are based on the original TSI data and their degradation corrections including knowledge of specific instrumental issues, these new composites heavily rely on statistical analysis and modelling of the original instrumental records.
These methods are good to close gaps in the data, whereas 
it is not clear to what extent such statistical modelling can recover hidden long-term variability (if any).
One advantage is, however, that the composite by \citet{dudok_de_wit_methodology_2017} was the first to supply uncertainty estimates (shown as shaded area in Fig. \ref{fig:tsicomposites}).
Uncertainty is formally also provided by \citet{montillet_data_2022}. However, it only includes the statistical and methodological errors and ignores the real uncertainty of the data, thus it does not represent the true uncertainty in the long-term TSI changes and cannot provide any additional information on the potential magnitude of such changes. 
One more composite was presented by
\citet{gueymard_reevaluation_2018}, which however also relies on alternative activity proxies to combine individual instrumental series, as well as the PMOD and ACRIM composites. 
In this sense, this should rather be considered as a fusion of measurements and empirical modelling and cannot provide any independent information on the long-term variability.

The TSI composites mentioned above are shown in Fig.~\ref{fig:tsicomposites}, while Table~\ref{tab:irradiancecomposites} also lists periods covered by them and the minimum-to-minimum TSI change.
To allow a direct comparison of the longer-term changes, we smoothed the daily values by roughly 6 months (see figure caption for details) and offset all series in Fig.~\ref{fig:tsicomposites} such as to match the PMOD composite during the 2009 activity minimum.
The figure 
shows that whereas the variability on time scales up to the solar cycle is fairly consistent among the composites,
on longer time scales, they clearly diverge.
As has already been reported in earlier publications and reviews cited above, the PMOD composite and its update by \citet[][called PMOD-rev in Table~\ref{tab:irradiancecomposites}]{schmutz_changes_2021} show a steady weak decrease in TSI from the minimum in 1986 to the most recent ones in 2009 and 2019. 
The ROB TSI first decreased between the minima in 1996 and 2009, and then slightly increased towards 2019. All changes are, however, statistically insignificant, being in the range of -3.9--2.6 $\times 10^{-3} $Wm$^{-2}$y$^{-1}$ \citep[see also][]{dewitte_total_2016}. 
The ACRIM composite including the version by \citet[][called ACRIM-rev in Table~\ref{tab:irradiancecomposites}]{scafetta_empirical_2023} shows a significant increase from 1986 to 1996, which is attributed to the sudden change in the sensitivity of the {\em Nimbus-7} Earth Radiation Budget (ERB) radiometer in 1989 \citep{hoyt_nimbus_1992,lee_iii_long-term_1995,frohlich_solar_2006,krivova_acrim-gap_2009}.
According to this composite, TSI then significantly decreased from 1996 to 2009 and then further to 2019 resulting overall in a very slightly increasing trend of 1.8$\times$ $10^{-3}$ Wm$^{-2}$y$^{-1}$ over the entire period of direct measurements. 
The statistically-constructed composite by \citet{dudok_de_wit_methodology_2017} assigns each instrument a weight based on a statistical estimate of its uncertainty.
The authors apply some corrections to ACRIM1, ACRIM2, HF, and ERBE data following \citet{frohlich_solar_2006} and then average the weighted data from the various instruments.
Nevertheless, the {\em Nimbus-7} ERB jump affects this composite as well, resulting in a TSI increase from 1986 to 1996, even though a roughly factor of three lower than in the ACRIM composite. Between 1996 and 2009, it shows a decrease, as all other series do. 
The new statistical composite by \citet{montillet_data_2022} using a data fusion approach \citep{cocchi_data_2019} suggests a steady TSI decrease over the whole period covered by observations.
Figure \ref{fig:tsicomposites} also shows the uncertainty in the composites as estimated by \citet{dudok_de_wit_methodology_2017}.
Unsurprisingly, the uncertainty rises when going back in time from about 0.1 Wm$^{-2}$ after 2000 to about 0.5 Wm$^{-2}$ towards the beginning of the monitoring.
It is evident also that most of the differences between the various produced composites lie within the $1\sigma$ uncertainty as estimated by \citet{dudok_de_wit_methodology_2017}.
Also, \citet{montillet_data_2022} conclude that the potential long-term change in TSI cannot be identified in the available data.

\begin{table*}
	\caption{List of TSI composites. Listed are the acronym of the composite (if existing), relevant publication, years covered by the composite, and the minimum to minimum trends between cycles 22 and 23 (1986--1996), 23 and 24 (1996--2009), 24 and 25 (2009--2019), 22 and 24 (1986--2009), as well as 22 and 25 (1986--2019).
	}
	\label{tab:irradiancecomposites}     
	\centering                      
	\begin{tabular}{ll*{6}{c}}       
\hline
		Acronym & Reference &  Years &\multicolumn{5}{c}{Trend between indicated activity minima [Wm$^{-2}$y$^{-1}$]$\times10^{-3}$} \\
		& &  & 22--23& 23--24& 24--25& 22--24& 22--25\\
\hline
PMOD &\cite{frohlich_solar_2006}    	  &1978--2018& -1.6 & -10.0 &    - &-6.5 &-\\
PMOD-rev&\cite{schmutz_changes_2021}		  &1978--2020& -3.3 &  -8.5 & -2.9 &-6.4 &-5.2\\[2mm]
ROB &\cite{dewitte_centennial_2022}		  &1991--2023&  - &  -3.9 &  2.6 & - & -\\[2mm]
ACRIM&\cite{willson_composite_2003} 	  &1978--2013& 48.0 & -24.9 &   -  & 7.6 & -\\
ACRIM-rev  &\cite{scafetta_empirical_2023}		  &1981--2022& 51.0 & -24.2 &   -14.5  & 10 &1.8\\[2mm]
\multicolumn{8}{c}{Statistical models}\\
    &\cite{dudok_de_wit_methodology_2017}&1978--2015&17.7 &  -12.5 & - & 0.9 & -\\
CPMDF&\cite{montillet_data_2022}		  &1978--2023& 1.0 & -15.9 & -5.9 &-8.4 &-7.6\\
\hline
	\end{tabular}
\end{table*}

\subsection{Magnetic field}

\begin{figure*}[]
	\centering
	\includegraphics[width=0.95\linewidth]{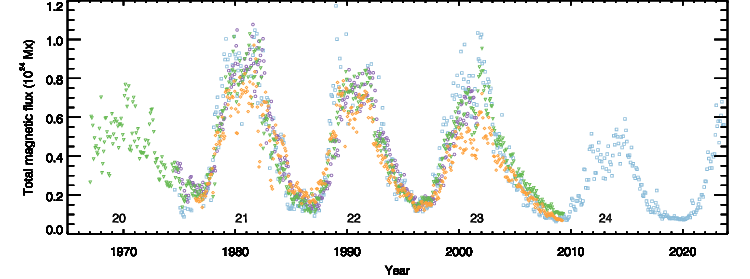}
	\caption{Evolution of the solar total photospheric magnetic flux. Data shown are from Wilcox Solar Observatory (WSO, orange rhombuses), National Solar Observatory at Kitt Peak (NSO, purple circles), and Mount Wilson Observatory (MWO, green triangles), as well as Kitt Peak Vacuum Telescope, SOHO/MDI, and SDO/HMI (light blue squares), cross-calibrated by \citet{yeo_analysis_2014}. WSO, NSO, and MWO are computed from Carrington maps, while the others are from daily full-disc magnetograms  \citep[here shown as monthly means multiplied by four to bring them roughly to the same scale as the other data;][]{riley_multi-observatory_2014}.
	}
	\label{fig:magflux}
\end{figure*}

The variability of the solar quantities relevant to climate studies (see Sect.~\ref{sec:intro}) is brought about by the Sun's incessantly evolving surface magnetic field. It is therefore of interest to look at how the surface magnetic field has varied in the past. More or less regular observations are available since the late 1960s \citep{howard_mount_1976,livingston_kitt_1976}, that is only slightly longer than irradiance measurements. The available ground- and space-based data are summarised in Fig.~\ref{fig:magflux} and suggest a continuous downward trend in the total magnetic flux during quiet periods from cycle 20 to cycle 24. Also the amplitude of the solar cycle variability has been declining from cycle 21 to cycle 24. 
The overall trend is consistent with those in the PMOD, \citet{schmutz_changes_2021} and \citet{montillet_data_2022} TSI composites, while the disagreement with the increasing trend of ACRIM between 1986 and 1996 is remarkable.

Figure~\ref{fig:magflux} also shows that solar cycle 25 is at a very similar level to or has surpassed the amplitude of cycle 24 even though cycle 25 is still in its ascending phase and has not yet reached its maximum. This is putting to rest claims about the Sun having entered a grand solar minimum state after cycles 23--24 \citep[see also Fig.~\ref{fig:tsicomposites} for the measured TSI values over the same period and ][for sunspot records]{clette_recalibration_2023,peguero_critical_2023}.

\section{Reconstructions of past solar activity and irradiance}
\label{se:recs}

Space-based measurements of TSI are invaluable for solar activity and irradiance studies.
Nevertheless, covering just slightly over the four most recent decades, they are clearly insufficient to allow reliable inferences about solar influence on Earth's climate over longer periods.
More so, as the period covered by observations has been heavily affected by human activities \citep[see][]{masson-delmotte_annex_2021,richardson_erroneous_2022}.
Reconstructions of past solar activity require suitable models and proxies.

\subsection{Irradiance models and reconstructions} 
\label{sec:tsimodels}

\begin{figure*}[]
	\centering
		\includegraphics[width=0.95\linewidth]{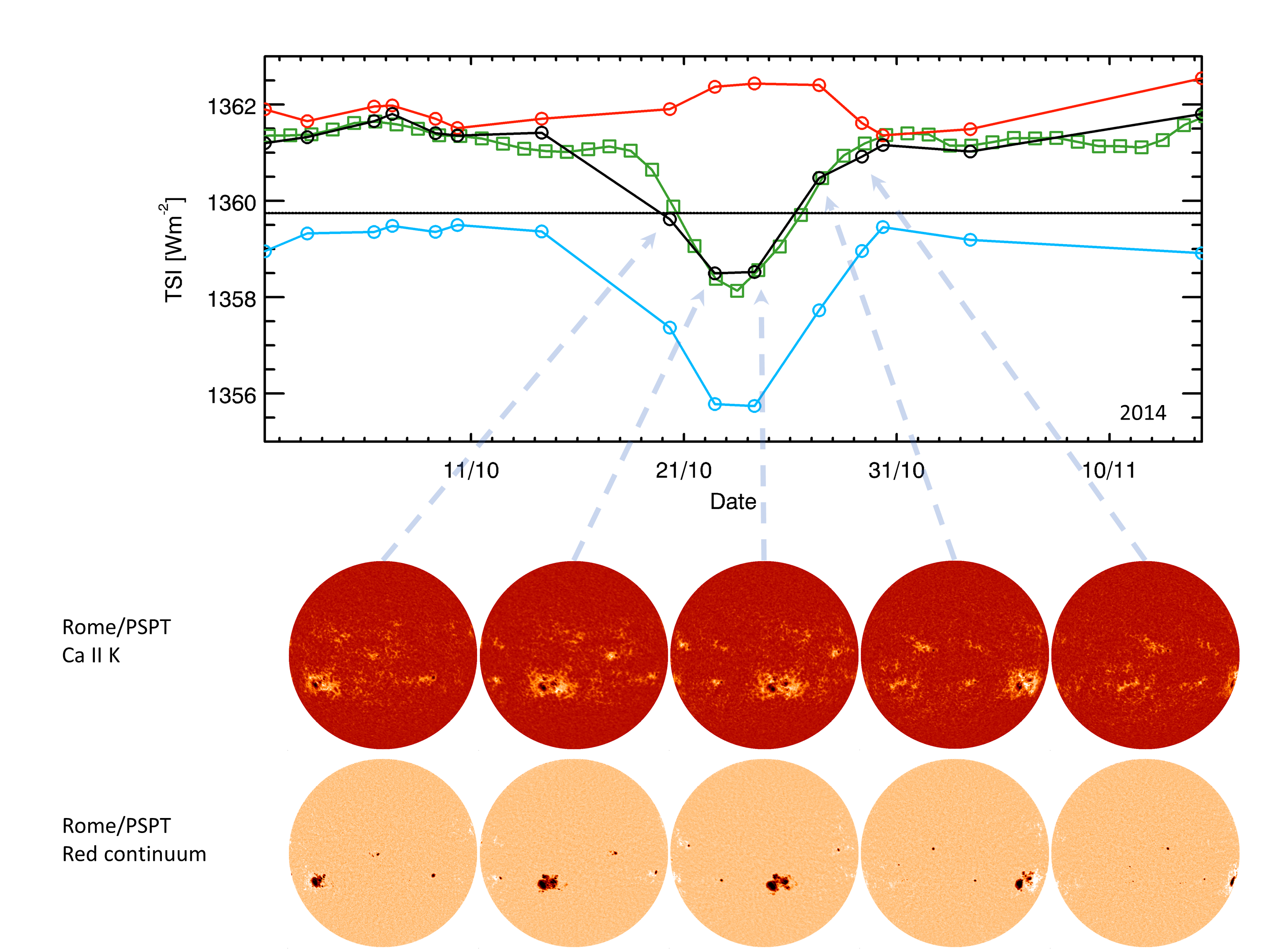}
	  \caption{ \textit{Top:} Daily TSI values over a 6-week interval starting from 1st of October 2014 (around the maximum of solar cycle 24). 
	  We show the values from the PMOD TSI composite (green squares) and the reconstruction with the SATIRE model from  Rome/PSPT \ca observations (black circles). 
	  Also shown are the individual contributions of faculae (red) and sunspots (ciel), as deviations from the mean level of TSI at 1359.7 Wm$^{-2}$ (horizontal black line). 
	  \textit{Bottom:} Full-disc Rome/PSPT observations in the \ca line (upper row) and in a red continuum band (lower row) on 5 days during the interval shown in the upper panel. The images have been compensated for limb-darkening with the method by \citet{chatzistergos_analysis_2018} and oriented to show the solar north pole at the top.}\label{fig:faccontribution}
\end{figure*}

Already soon after the first space-based irradiance measurements revealed its variable nature, it was recognised that the solar surface magnetic field played an important role in this variability, with sunspots leading to depressions in TSI \citep[e.g.][]{willson_observations_1981} and faculae together with network regions causing enhancements in TSI \citep[e.g.][and references therein]{oster_solar_1982,shapiro_are_2016}.
Figure \ref{fig:faccontribution} shows the TSI as well as individual contributions to its changes from sunspots and faculae computed with the Spectral And Total Irradiance REconstruction \citep[SATIRE;][]{krivova_reconstruction_2003} model employing \ca observations (see Sect.~\ref{sec:caiik}, \citealp{chatzistergos_reconstructing_2021-1})
over a 6-week interval around the maximum of solar activity of cycle 24.
Subsequent progress in modelling techniques has confirmed this link and
more recent studies have unambiguously shown that the evolution of the solar surface magnetic field accounts for basically all (within the uncertainties of measurements; see Fig. \ref{fig:tsicomposites} and  \citealt{dudok_de_wit_methodology_2017,montillet_data_2022}) TSI variations on timescales of days to decades \citep{krivova_reconstruction_2003,shapiro_nature_2017,yeo17b}, thus leaving essentially no room for other mechanisms.

\begin{table*}
	\caption{Studies that reconstructed solar irradiance back to at least 1750. Listed are the model acronym (if existing), reference to the latest version, input data used for the reconstruction, full period covered, and the TSI difference between 1986 and 1700. The subsequent lines within each individual box list previous versions of the same models, if exist, however, we list the TSI differences only for the most recent versions of the models. The models are grouped into empirical (top part of the table) and semi-empirical (bottom part) ones. Marked with grey are the series which at least at some point lie outside the constraint imposed by \cite{yeo20b}. }
	\label{tab:irradiancereconstructions}     
	\centering               
	\small
	\begin{tabular}{lll*{3}{c}}       
\hline
		Model & Study &  Data& Years &TSI$_{1986}$ - TSI$_{1700}$ \\
		 & & &  & [Wm$^{-2}$]\\
		\hline\hline\\[-8pt]       
		\multicolumn{5}{c}{Empirical models}\\
		\hline\hline\\[2pt]       
				&\cite{mordvinov_reconstruction_2004}&HoSc98 GSN& 1610--2003&$\sim$1.4\\ 
				\hline
				&\cite{svalgaard_no_2007}&SvSc16 GSN& 1610--2017&-0.09\\
				\hline
  				&\cite{foukal_new_2012}&HoSc98 GSN+\cite{foukal_behavior_1996} \ca plage areas&1610--2009&0.09\\
				\hline
  				&\cite{tapping_solar_2007}&ISNv1+ HoSc98 GSN&1600--2006&$\sim$1.00\\
				\hline
 				&\cite{foster_reconstruction_2004}&ISNv1&1700--1996&$\sim$1.23\\ 
 				\hline
 				&\cite{schatten_solar_1990}&ISNv1&1610--1993&$\sim$0.14\\
				\hline
				&\cite{wang21}&ISNv2&1700--2019&0.2\\
				&\cite{wang05}&&&\\
		\hline
				&\cite{dewitte_centennial_2022}&ISNv2&1700--2020&0.04\\
		\hline
				&\cite{privalsky_new_2018}&ISNv2&1750--1978&-\\
		\hline
				&\cite{penza_total_2022}&MEA20+CEA20&1513--2001&2.2\\
				\hline
            &\cite{hoyt_discussion_1993}&Sunspot indices&1700--1992&\cellcolor{darkergrey}{4.0}\\
		\hline
			&\cite{georgieva_reconstruction_2015}&aa-index&1699--2008&$\sim$1.6\\
		\hline
		NRLTSI &\cite{lean_estimating_2018}&Cosmogenic isotopes+GSN&850--2016&0.5\\
		&\cite{coddington_solar_2016}\\
		&\multicolumn{3}{l}{\cite{lean_evolution_2000,lean_estimating_1992,lean_reconstruction_1995}}&\\
				\hline
				&\cite{mordvinov_reconstruction_2004}&Cosmogenic isotopes& 850--2003&$\sim$1.6--1.7\\
				\hline
				&\cite{delaygue_antarctic_2011}&Cosmogenic isotopes&850--1961&-\\
				&\cite{bard_solar_2000,bard_comment_2007}\\
				\hline
				&\cite{steinhilber_9400_2012}&Cosmogenic isotopes&-7400--1988&1.2\\
				&\cite{steinhilber_total_2009}\\
				\hline
 				&\cite{roth_reconstruction_2013}&Cosmogenic isotopes&-8050--2005&0.96\\ 
				\hline
 				&\cite{foster_reconstruction_2004}&Cosmogenic isotopes&1424--1989&$\sim$0.76\\
				\hline
 				&\cite{fedorov_role_2021}&Cosmogenic isotopes&-7290--2019$^a$&1.4\\
				\hline
				&\cite{velasco_herrera_reconstruction_2015}&HoSc98 GSN + Machine learning extrapolation&1000--2109&$\sim$0.17--1\\
				\hline
 				&\cite{abduallah_reconstruction_2021}&Machine learning extrapolation&-6755--1885&-\\
\\[4pt]  		\hline
		\hline         
		\multicolumn{5}{c}{Semi-empirical models}\\
		\hline\hline \\[2pt]          
		\hline
		 &\cite{bolduc_Modelisation_2014}& RGO areas + HoSc98 GSN &1610--2013&  $\sim$0.6 \\ 
		\hline
		&\cite{fligge_solar_2000-1}&ISNv1&1700--1999&\cellcolor{darkergrey}{$\sim$4.3}\\
		&\cite{solanki_reconstruction_1999}\\
		\hline
		SATIRE-T &\cite{wu_solar_2018-2}& ISNv2+\cite{vaquero_level_2015}&1639--2017&  0.38 \\
		&\cite{krivova_reconstruction_2007,krivova_reconstruction_2010}\\
		\hline
		SATIRE-T2 &\cite{dasi-espuig_reconstruction_2016}& ISNv1+RGO areas&1700--2008&  0.79 \\
		&\cite{dasi-espuig_modelling_2014}\\
		\hline
		SATIRE-M &\cite{wu_solar_2018-2}& Cosmogenic isotopes &-6755--1885&  - \\
		&\cite{vieira_evolution_2011}\\
		\hline
		CHRONOS &\cite{egorova18}&Cosmogenic isotopes&-6000--2015&\cellcolor{darkergrey}{3.5--5.5}\\
		&\cite{shapiro11}&&&\\
\hline
	\end{tabular}
	\tablefoot{\textbf{Notes: }$^{\mathrm{a}}$ It is unclear how the series was produced for the period after 1950, while after personal communication the authors were not able to clarify this.}
\end{table*}

\begin{figure*}[]
	\centering
	\includegraphics[width=0.95\linewidth]{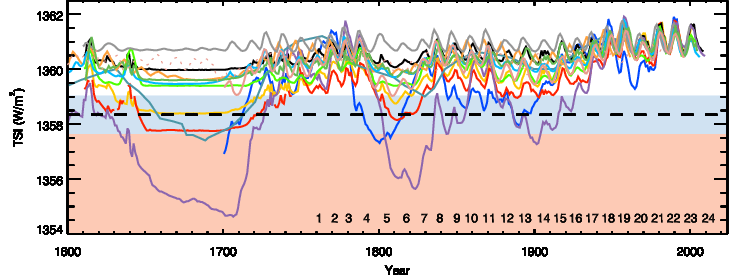}
	\includegraphics[width=0.95\linewidth]{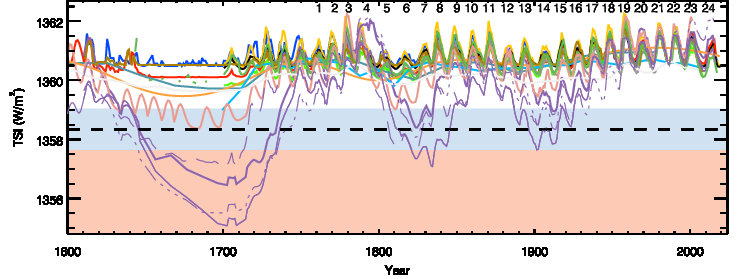}
	\caption{TSI reconstructions extending back to at least 1750. The two panels show models divided by the year of publication, in particular the top panel shows models published before 2012, while the bottom panel shows the currently most up to date models. 
		\textit{Top: } \citet[][grey]{schatten_solar_1990}, \citet[][blue]{hoyt_discussion_1993}, \citet[][NRLTSI; red]{lean_reconstruction_1995}, \citet[][yellow]{lean_evolution_2000}, \citet[][pale blue]{bard_solar_2000}, \citet[][bright green]{mordvinov_reconstruction_2004}, \citet[][pink, thick and thin lines for the reconstructions with concentrations of cosmogenic isotopes and the open solar flux by \citealt{solanki_evolution_2000} reconstructed from ISNv1, respectively]{foster_reconstruction_2004}, \citet[][black]{wang05}, 
		\citet[][SATIRE-T; green]{krivova_reconstruction_2007}, \citet[][ciel]{tapping_solar_2007}, \citet[][purple]{shapiro11}, and \citet[][orange]{delaygue_antarctic_2011}.
		\textit{Bottom: }
		\citet[][pale blue]{steinhilber_9400_2012}, \citet[][brown]{foukal_new_2012}, \citet[][light grey]{roth_reconstruction_2013}, 		\citet[][ciel]{georgieva_reconstruction_2015}, \citet[][SATIRE-T2; bright green]{dasi-espuig_reconstruction_2016}, \citet[][SATIRE-T; green]{wu_solar_2018-2}, \citet[][CHRONOS; purple, shown are four different versions of this model based on the modulation potential used; thick solid line for \citealt{usoskin_solar_2016}, thin solid line for McCracken \& Beer 2017, dashed line for \citealt{shapiro11}, and sparse dashed line for \citealt{muscheler_revised_2016}]{egorova18}, \citet[][NRLTSI; red]{lean_estimating_2018}, \citet[][grey]{privalsky_new_2018}, \citet[][updated in 2018; blue]{svalgaard_no_2007}, \citet[][black]{wang21}, \citet[][orange]{fedorov_role_2021}, \citet[][yellow]{dewitte_centennial_2022}, and \citet[][pink]{penza_total_2022}. The horizontal black dashed line marks the dimmest state of the Sun that could be reached if the global dynamo ceased \citep[][see Sect.~\ref{sec:dimmeststate}]{yeo20b}. The uncertainty range of this constraint is shown by light blue shading, while red shading marks values clearly outside the allowed range for a possible TSI level during a deep grand minimum of solar activity.
		All series are offset to match the value of \cite{montillet_data_2022} composite over 1986.  Exceptions are for the series by \cite{steinhilber_9400_2012,bard_solar_2000,delaygue_antarctic_2011} and \cite{fedorov_role_2021}, which do not extend to 1986, for which we used their overlap with SATIRE-T over 1900--1980, after offsetting SATIRE-T to \cite{montillet_data_2022} composite over 1986.
		Shown are annual median values, while the numbers within each panel denote the conventional solar cycle numbering.}
	\label{fig:tsireconstructions}
\end{figure*}
\begin{figure*}[]
	\centering
	\includegraphics[width=0.95\linewidth]{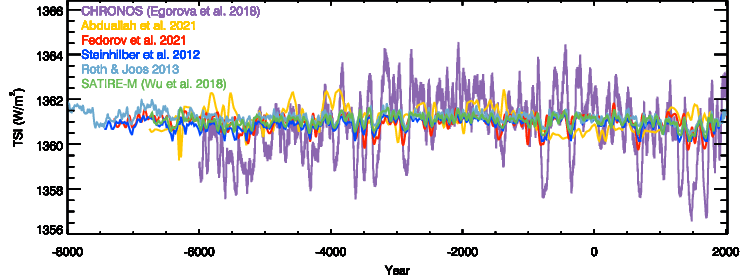}
	\caption{TSI reconstructions over the Holocene:  \citet[][SATIRE-M; green]{wu_solar_2018-2}, \citet[][blue]{steinhilber_9400_2012}, \citet[][light blue]{roth_reconstruction_2013}, \citet[][orange]{abduallah_reconstruction_2021}, \citet[][red]{fedorov_role_2021}, and \citet[][CHRONOS; purple]{egorova18}. All series were offset to match over the period 1850--1950. }\label{fig:tsireconstructions_cosmogenic}
\end{figure*}
\begin{figure*}[]
	\centering
	\includegraphics[width=0.95\linewidth,trim={0 0.8cm 0cm 0.1cm},clip]{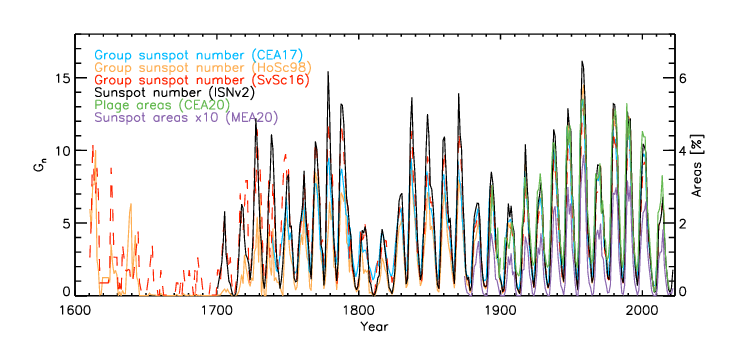} 
	\includegraphics[width=0.95\linewidth,trim={0 0cm 0cm 0.55cm},clip]{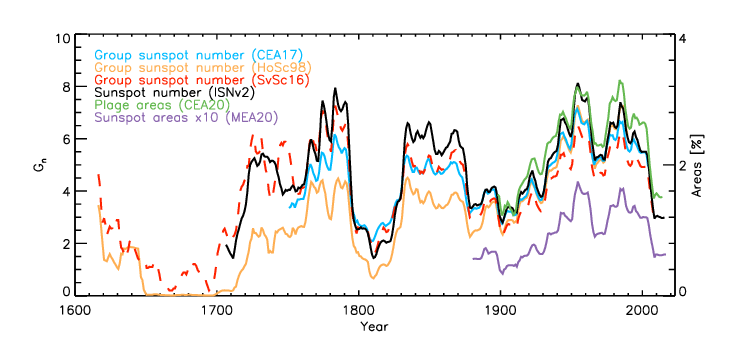}
	\caption{Selected indices of solar activity extending back to at least 1900. Top panel shows annual values while the bottom panel shows 11-year running means. Shown are the group sunspot number, $G_n$, by \citet[][HoSc98, orange]{hoyt_group_1998}, \citet[][SvSc16, red]{svalgaard_reconstruction_2016}, and \citet[][CEA17, ciel]{chatzistergos_new_2017} as well as the International sunspot number v2 from \citet[][ISNv2, black]{clette_new_2016-1}, the \ca plage area composite by \citet[][CEA20, green]{chatzistergos_analysis_2020}, and the sunspot area by \citet[][MEA20, purple]{mandal_sunspot_2020}. Shown are projected areas as fractions of the visible solar disc (see the ordinate axis on the right-hand side for both). The sunspot areas have been multiplied by 10 to bring them closer to the level of the plage areas, while the ISNv2 series was divided by 20.13 
	to bring it roughy to the same level as the group sunspot number records. 
	}
	\label{fig:solarindices}
\end{figure*}

\begin{figure*}[]
	\centering
		\includegraphics[width=0.95\linewidth]{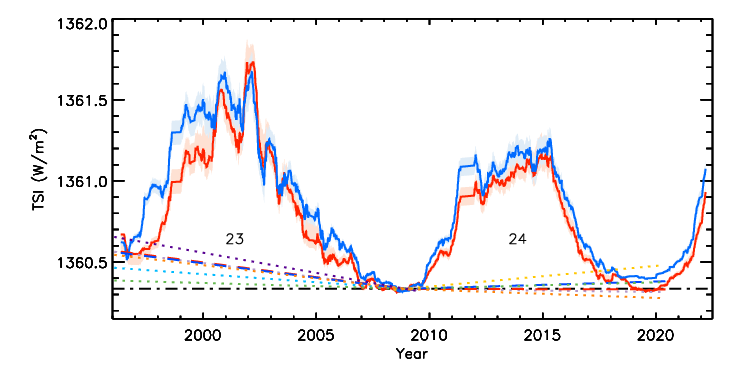}
	\caption{TSI variations reconstructed from Rome/PSPT \ca data using the \citet[][red]{chatzistergos_modelling_2020} and \citet[][blue]{chatzistergos_reconstructing_2021-1} models. The shaded surfaces encompass the reconstructions by using different reference series.
	The straight dotted lines are linear fits between subsequent minima as obtained from the PMOD (light blue), ROB (green), ACRIM (purple), \citet[][orange]{montillet_data_2022}, 
	SORCE/TIM (yellow), and SATIRE-S \citep[][pink]{yeo_reconstruction_2014} TSI series, while the straight dashed lines are for the reconstructions with \ca data by \citet[][red]{chatzistergos_modelling_2020} and \citet[][blue]{chatzistergos_reconstructing_2021-1}. The black dot-dashed line denotes 0-trend line. 
	The numbers in the middle of the panel denote the conventional solar cycle numbering.}
	\label{fig:tsireconstructions_phsum}.

\end{figure*}

All currently existing models can be split into three main groups: purely empirical, semi-empirical, and physical.
Purely empirical models use indices of solar activity which are regressed to direct TSI measurements in order to reconstruct TSI variations \citep[e.g.][]{hudson_effects_1982,schatten_importance_1985,foukal_magnetic_1988,chapman_variations_1996,chapman_modeling_2013,preminger_photometric_2002,wang05,wang21,steinhilber_9400_2012,delaygue_antarctic_2011,zhao_suns_2012,coddington_solar_2016,tebabal_modeling_2015,yeo_empire_2017,privalsky_new_2018,mauceri_neural_2019,chatzistergos_modelling_2020}.
Some empirical models also include machine learning extrapolations \citep[e.g.][]{velasco_herrera_reconstruction_2015,abduallah_reconstruction_2021}, whose performance is, however, questioned by the stochastic nature of solar activity \citep[see e.g.][and references therein]{cameron_solar_2019,charbonneau20,petrovay_solar_2020}.
Semi-empirical models rely on solar observations to determine the location and area covered by magnetic features, while solar model atmospheres and spectral synthesis codes are used to calculate the brightness spectra of these features \citep[e.g.][]{fligge_model_1998,fligge_solar_2000-1,fligge_modelling_2000-1,solanki_solar_1998,solanki_reconstruction_1999,ermolli_modelling_2001,ermolli_modeling_2003,ermolli_recent_2011,krivova_reconstruction_2003,krivova_reconstruction_2007,krivova_reconstruction_2009,krivova_reconstruction_2010,krivova_solar_2011,penza_modeling_2003,fontenla_semiempirical_2006,fontenla_high-resolution_2011,wenzler_reconstruction_2006,crouch_model_2008,haberreiter_nlte_2008,haberreiter_reconstruction_2014,shapiro11,vieira_evolution_2011,ball_reconstruction_2012,bolduc_reconstruction_2014,dasi-espuig_modelling_2014,dasi-espuig_reconstruction_2016,yeo_reconstruction_2014,wu_solar_2018-2,chatzistergos_reconstructing_2021-1}. 
Both empirical and semi-empirical models have free parameters that need to be set by comparison to direct TSI data.
The number of free parameters varies among the models.
Finally, physical models do not rely on TSI measurements for their reconstructions at all and compute the irradiance variations from 3D magnetohydrodynamics (MHD) simulations of the solar atmosphere \citep{yeo17b}.
The model by \citet{yeo17b} is the first, and to the best of our knowledge the only to date, model of this type, see also Sect.~\ref{sec:dimmeststate}.

Figure~\ref{fig:tsireconstructions} shows various TSI reconstructions extending back to at least 1700s.
Some of these models have regularly been revised, refined, and updated.
Nevertheless, often older versions are used in the literature ignoring the fact that newer versions, overriding the earlier ones, are available.
We caution against using outdated reconstructions. 
In the top panel of the figure we show the older reconstructions (roughly older than 10 years), while the bottom panel shows more recent ones.
Interestingly, the more recent reconstructions show a somewhat better consistency among them than the older ones.
However, the disagreement remains. Thereby, various reconstructions show a lot of similarities on time scales of the solar cycle and less, while most of the disagreement now concerns the amplitude of the secular variability.
Table \ref{tab:irradiancereconstructions} lists some key characteristics of TSI reconstructions that extend back to at least 1750.
In this table we have grouped studies that used the same model together but provide a TSI difference between 1700 and 1986 only for the latest version of each model.
Most series (21 out of 24 models from Table \ref{tab:irradiancereconstructions} covering 1700 to 1986) suggest a TSI difference between 1700 and 1986  of about -0.09--2.2 Wm$^{-2}$, with only two older models by \citet{hoyt_discussion_1993} and  \citet[][this model should be, however, considered replaced by the newer SATIRE-family models]{fligge_solar_2000-1}, as well as that by \citet{egorova18} favouring a higher level of variability (see also Sect.~\ref{sec:dimmeststate}) of 3.5--5.5 Wm$^{-2}$. 
However, we note that the secular trend is not an intrinsic characteristic of all of the  TSI reconstructions listed in Table \ref{tab:irradiancereconstructions}, since on many of them, especially the older ones \citep[e.g.][]{lean_estimating_1992,lean_reconstruction_1995,hoyt_discussion_1993,solanki_reconstruction_1999,fligge_solar_2000-1,bard_solar_2000,bard_comment_2007}, it was rather imposed based on information on the magnitude of variability found for Sun-like stars. These old studies relied heavily on many assumptions which were shown to be incorrect \citep[e.g.,][]{hall_chromospheric_2004,hall_activity_2009,wright_we_2004} and thus old TSI reconstructions using them to impose the secular trend should not be used.

Figure \ref{fig:tsireconstructions_cosmogenic} shows some of the more recent TSI reconstructions extending back to at least 6000 BCE.
Again, most of them show comparatively weak secular variations in TSI over the entire Holocene, and only the reconstruction by \citet[][]{egorova18} suggests higher variability.

\subsection{Past activity proxies}
\subsubsection{An overview}
\label{sec:proxy_overview}

Despite the differences in the model methodologies, they all require appropriate data to describe emergence of the solar surface magnetic field in the past.
Full-disc observations of magnetic regions generally return more accurate reconstructions than disc-integrated indices. This is because knowing the positions of the various surface on the visible disc allows taking the dependence of their brightness on the limb distance (centre-to-limb variation) into account \citep[see, e.g.,][]{ortiz_intensity_2002,yeo_intensity_2013}.
Such data (magnetograms or images at various wavelengths), however, only exist over roughly the same period as the irradiance measurements.
The only exceptions are areas and positions of sunspots (see Sect.~\ref{sec:spots}) since 1874, often used in reconstructions, and \ca observations since 1892 (Sect.~\ref{sec:caiik}), until now barely used for this purpose.

The longest direct and most widely used proxy of solar activity is that of sunspot counting, the sunspot number, going back, although with deteriorating quality, to early 17th century (see Sect.~\ref{sec:spots}).
Since 1874 also records of sunspot areas and positions are available.

Sunspots are the strongest concentrations of the surface magnetic field on the Sun.
However, most of the magnetic flux emerges in smaller regions \citep[see][]{harvey93a,thornton11,krivova_modelling_2021}, which appear bright.
These smaller regions, called faculae (or plage if observed in the chromosphere, that is in the level of the solar atmosphere lying higher than the photosphere where spots are observed) and network, appear bright.
These bright features are the main driver of changes in solar irradiance on time scales of years and longer.
At the same time, their evolution with time is not perfectly represented by the evolution of sunspots.
For example, over the solar cycle, the emergence rate of stronger magnetic features varies significantly more than that of weaker features \citep{harvey93a}.
More critically, during quiet periods, like solar activity minima, when no sunspots emerge on the surface of the Sun, the magnetic field continues emerging in smaller concentrations.
However, while the number of observed sunspots is zero, their emergence does not carry information on the emergence of weaker regions during such periods. This becomes particularly critical \citep[see][]{krivova_modelling_2021} over extended periods of reduced solar activity, the so-called grand minima, such as the well-known Maunder minimum.
Thus, the absence, until recently, of a reliable proxy of facular/plage emergence in the past is the main reason of the existing uncertainty in secular irradiance variability (see Sect.~\ref{sec:tsimodels}), since models had to rely on indirect proxies or involve questionable assumptions.

As already mentioned above, full disc magnetograms provide excellent and most straightforward information on facular regions. However, regular high-quality data exist since 1970s \citep{livingston_kitt_1976} and thus cover essentially the same period as direct TSI measurements.
Besides magnetograms, there are several disc-integrated chromospheric indices \citep{ermolli_solar_2015} that can provide information on facular regions and have been used by various models, such as the 10.7 cm radio flux \citep[available since 1947;][]{tapping_next_2013}, Lyman $\alpha$ emission \citep[available since 1969;][]{woods_improved_2000}, Mg II index \citep[available since 1978;][]{heath_mg_1986,snow_comparison_2014}, He~{\sc i} (10830\AA) equivalent width (since 1977); Ca~{\sc ii}~(8542\AA) central depth (since 1978), H$\alpha$ central depth (since 1984), CN (3883~\AA) bandhead index \citep[since 1979;][]{livingston_sun-as--star_2007}, and \ca 1~\AA~ disc-integrated emission index \citep[since 1976;][]{bertello_ca_2017}. 
All these disc-integrated series cover either shorter periods than the direct TSI measurements or marginally longer, with F10.7 being the longest one (since 1947).
To address this problem, a lot of effort has been recently put into exploitation of the exclusive potential of historical solar \ca observations \citep[see e.g.][and Sect.~\ref{sec:caiik}]{chatzistergos_analysis_2017,chatzistergos_full-disc_2022}.

Another proxy of solar magnetic activity, also used for irradiance reconstructions, in particular on time scales of millennia, is the concentration of cosmogenic isotopes in terrestrial archives, such as ice sheets or ancient plants.
Cosmogenic isotopes are produced by the highly energetic particles of the GCRs (and also to a much lesser extent by SEPs, so that only a few most extreme solar events could be identified in radionucide data, \citealt{cliver_extreme_2022}).
The flux of GCRs is modulated by the solar open magnetic flux, and so the production rates and eventually the concentrations of the cosmogenic isotopes in terrestrial archives ($^{10}$Be in ice sheets and $^{14}$C in ancient plants) are linked to solar magnetic activity.
More details can be found in \citet[][]{usoskin_history_2023}.

Cosmogenic isotope data are, however, only an indirect proxy and do not allow distinguishing between sunspots and faculae without further assumptions or models. Reconstructions based on such data are thus either scaled to direct irradiance measurements, which as we showed in Sect.~\ref{sec:meas} do not provide a unique information on the secular trend, or to other, e.g., sunspot-based, reconstructions.
Another limitation of radionuclide-driven reconstructions is their low temporal resolution.
Typically only decadal data are available over the Holocene \citep[although
see][]{beer_use_1990}. 
Recently high-quality annual $^{14}$C data over the last Millennium became available \citep{brehm_eleven-year_2021,kaiser_kudsk_solar_2022}.
\citet{usoskin_solar_2021} have used these data to reconstruct the yearly sunspot number over the period 971–-1900, which covers 85 full solar cycles.
The mean length of the cycles over this period was found to be 10.8~years.
This reconstruction has significantly increased the number of known solar cycles, from 36 cycles since 1610 covered by direct observations (including four essentially unresolvable cycles during the Maunder minimum) to 96 cycles all together.
The considered period covered five grand minima of solar activity, with a total duration of about 430 years. These include the Oort, Wolf, Sp\"orer, Maunder and Dalton minima, although Oort and Dalton minima were short and not as deep as the other three grand minima.

\subsubsection{Sunspot records}
\label{sec:spots}

Direct data providing information on sunspots go back to the early 17th century.
These are expressed as either combined countings of sunspots and sunspot groups, called the international sunspot number \citep[ISNv1;][]{clette_wolf_2007}, or the number of sunspot groups (GSN) alone as introduced by \citet[][HoSc98, hereafter]{hoyt_group_1998}.
The last decade was marked by a rekindled interest in and extensive study of sunspot data, leading to a recovery of many new data, as well as corrections of existing records \citep[e.g.][]{arlt_sunspot_2013,arlt_sunspot_2016,vaquero_revised_2016,carrasco_sunspot_2018-1,carrasco_two_2019,carrasco_note_2021,carrasco_forgotten_2021,hayakawa_daniel_2021,hayakawa_overview_2022,vokhmyanin_sunspot_2020,bhattacharya_modern_2021}.
Furthermore, various issues with the cross-calibration and compilation approaches of records of individual observers have been realised \citep{clette_new_2016-1,clette_new_2018,lockwood_tests_2016-2,usoskin_dependence_2016}, leading to new techniques and several alternative GSN series \citep{usoskin_new_2016,usoskin_robustness_2021,willamo_updated_2017,chatzistergos_new_2017,svalgaard_reconstruction_2016,cliver_discontinuity_2016} as well as new versions of ISN \citep[ISNv2--ISNv2.3,][]{clette_new_2016-1,bhattacharya_rudolf_2023,bhattacharya_scale_2023}.
Compared to HoSc98, the new GSN series have a higher amplitude prior to the 20th century, with the 
series by \citet{svalgaard_reconstruction_2016,cliver_discontinuity_2016} being significantly higher and the series by \citet{chatzistergos_new_2017,usoskin_robustness_2021} lying between the latter and HoSc98 (see Fig.~\ref{fig:solarindices}).
Comparisons to other proxies (such as, for example, $^{14}$C,  \citealp{usoskin_solar_2021}, but also some other data, see \citealt{clette_recalibration_2023} for further details) support the new revisions showing moderate to high levels of activity over the 18th and 19th centuries \citep[e.g.][]{chatzistergos_new_2017, usoskin_robustness_2021}.
A comparison to $^{44}$Ti \citep{asvestari_assessment_2017} disfavours high levels of activity over 18th and 19th centuries, while moderate levels of activity are consistent with this proxy, too.

Since 1874 also other parameters of sunspots have been recorded, such as their areas and positions on the solar disc.
These provide a more accurate information on solar activity, particularly important for irradiance reconstructions, as the amount of radiation blocked by sunspots depends on their size and heliospheric angle (sunspots closer to the disc centre are darker than those closer to the limb).
Spatial patterns of sunspot emergence are also important for understanding the solar dynamo action.
A record of sunspot area measurements has been kept at the Royal Greenwich Observatory between 1874 and 1976 \citep{willis_greenwich_2016}.
Due to systematic differences between the measurements by various other sites, a cross-calibration of the data is needed to avoid fake trends when using the data for understanding past solar activity.
The most recent and accurate cross-calibration is offered by \citet[][shown in Fig. \ref{fig:solarindices}]{mandal_sunspot_2020}.

\subsubsection{\ca observations}
\label{sec:caiik}

\ca data are full-disc observations of the Sun in the singly ionised Ca line at 393.37 nm \citep{chatzistergos_analysis_2017,chatzistergos_full-disc_2022,ermolli_potential_2018} (see Fig. \ref{fig:faccontribution} for some examples of modern \ca observations).
We note that historical full-disc photographic observations going back to 1874 are also available in white-light \citep[][]{foukal_curious_1993}.
Their use is, however, restricted due to the fact that white-light data show facular regions essentially only near the limb, thus not describing accurately the entire facular contribution to TSI.
The value of \ca data for past solar activity and irradiance studies comes from their link to solar surface magnetism \citep[][and references therein]{chatzistergos_recovering_2019}.
The earliest \ca observations were made in 1892, and monitoring of the Sun in \ca line continues to this day at a number of sites around the globe \citep{chatzistergos_full-disc_2022}.
These data are a real treasure trove for past solar activity studies since they can be used to recover information on solar surface magnetism over periods for which direct measurements do not exist.

Unfortunately, \ca data suffer from various issues rendering their application  not straightforward.
A critical aspect is that historical data have been stored on photographic plates, which are not linear detectors. 
The images are also affected by various artefacts which have arisen at various stages of their life \citep[see, e.g.,][]{chatzistergos_analysis_2017,chatzistergos_full-disc_2022}.
For instance, there are brightness variations across linear (or curved) segments due to problems during the observation, there are smudges due to the accumulation of dirt during the storage of the images, or artefacts introduced during the digitisation of the plates  \citep[see, e.g.,][]{chatzistergos_full-disc_2022}.
All of these issues need to be accounted for before using \ca data for acquiring accurate results from them.

Because of the challenges of extracting accurate information from historical \ca data, most irradiance reconstructions from \ca data were restricted to using modern CCD-based images \citep{chapman_variations_1996,chapman_modeling_2013,walton_contribution_2003,ermolli_modeling_2003,ermolli_recent_2011,penza_modeling_2003,fontenla_bright_2018,puiu_modeling_2019,chatzistergos_modelling_2020,chatzistergos_reconstructing_2021}, which do not require photometric calibration and suffer from considerably fewer artefacts than the historical data.
These studies are particularly important to ascertain the ability of \ca data to be used for accurate irradiance reconstructions.
Since some CCD-based \ca archives, such as those from Rome/PSPT \citep[Precision Solar Photometric Telescope;][]{ermolli_rome_2022} and San Fernando \citep{chapman_solar_1997}, have long and consistent observations they can also be used to assess the cycle minimum to minimum trend in TSI, thus avoiding degradation effects of satellite measurements or cross-calibration issues of different direct TSI series.
This is because the data used for these reconstructions are independent of the TSI measurements, while output of the models such as SATIRE or the one by \citet{chatzistergos_modelling_2020} is only very little affected by the choice of the reference TSI record used to constrain the free parameters \citep[e.g.][]{chatzistergos_reconstructing_2021}.
\citet{chatzistergos_modelling_2020} used Rome/PSPT data to reconstruct solar irradiance variations back to 1996 with an empirical model based on the integrated intensity in \ca and a continuum interval at the blue part of the spectrum (409.2 nm), following the approach by \citet{chapman_comparison_2012,chapman_modeling_2013}.
They reported a generally decreasing TSI trend since 1996. 

Figure \ref{fig:tsireconstructions_phsum} shows the activity minimum-to-minimum changes for various direct TSI series and reconstructions, including also the TSI reconstructions from \ca data using the \citet{chatzistergos_modelling_2020} and adapted SATIRE \citep{chatzistergos_reconstructing_2021-1,chatzistergos_reconstructing_2021} models.
Both TSI reconstructions were updated with newer data from Rome/PSPT as well as considering some corrections for the early data, while the reconstruction following \citet{chatzistergos_reconstructing_2021-1} shown here 
also used the latest version of the \citet{mandal_sunspot_2020} database.
Both models return qualitatively the same trends, with values of $-18^{+5.1}_{-0.2}\times10^{-3}$ Wm$^{-2}$y$^{-1}$ and $-18^{+1}_{-1}\times10^{-3}$ Wm$^{-2}$y$^{-1}$ between the 1996 and 2008 minima for the empirical reconstruction by \citet{chatzistergos_modelling_2020} and the semi-empirical Ca\,II-driven SATIRE \citep{chatzistergos_reconstructing_2021}, respectively.
For the change between the 2008 and 2019 minima, we find $-0.39^{+0.03}_{-0.14}\times10^{-3}$ Wm$^{-2}$y$^{-1}$ and $3.8^{+0.03}_{-0.1}\times10^{-3}$ Wm$^{-2}$y$^{-1}$ for the reconstruction by \citet{chatzistergos_modelling_2020} and \citet{chatzistergos_reconstructing_2021-1}, respectively.

Use of photographic \ca data for irradiance reconstructions has been significantly more limited than that of CCD-based ones.
To our knowledge, the only TSI reconstructions from photographic \ca data are those by \citet{ambelu_estimation_2011}, \citet{foukal_new_2012}, \cite{xu_reconstruction_2021}, \cite{penza_total_2022}, and \citet{chatzistergos_reconstructing_2021-1}.
Thereby, \citet{ambelu_estimation_2011}, \citet{foukal_new_2012}, and \cite{xu_reconstruction_2021} used linear regression models to reconstruct TSI variations, which limits their reliability. 
Furthermore, these reconstructions were produced from photometrically uncalibrated \ca data and
used a single \ca archive without assessing the consistency of the series. Thus they are prone to inhomogeneities of the data \citep{chatzistergos_full-disc_2022,chatzistergos_analysis_2023}.
\cite{penza_total_2022} reconstructed TSI variations with an empirical model employing the plage area composite series by \cite{chatzistergos_analysis_2020}.

The only semi-empirical TSI reconstruction from photometrically calibrated historical \ca data is the one by \citet{chatzistergos_reconstructing_2021-1}.
Besides being able to photometrically calibrate the historical \ca data, their analysis benefited from more accurate processing techniques \citep{chatzistergos_analysis_2018,chatzistergos_analysis_2019,chatzistergos_delving_2019,chatzistergos_analysis_2020,chatzistergos_historical_2020} than in previous studies as well as from a considerably larger sample of \ca datasets.
The reconstruction by \citet{chatzistergos_reconstructing_2021-1} is currently limited to the period covered by direct TSI measurements, since the main goal of this first study was to set up the model and assess its performance using diverse \ca data.
This study has highlighted the potential of using \ca data to accurately recover TSI variations in the past provided the data are accurately and consistently processed \citep[such as with the methods by][]{chatzistergos_analysis_2018,chatzistergos_analysis_2019,chatzistergos_analysis_2020}. It also emphasised the importance of understanding and accounting for the different characteristics of the various available \ca data \citep{chatzistergos_full-disc_2022,murabito_invest_2023}.

\ca data have also been used to study the evolution of plage areas.
The first composite of plage areas from 43 available datasets was presented by \citet[][shown in Fig. \ref{fig:solarindices}]{chatzistergos_analysis_2020}.
\ca data are available since the late 19th century, significantly extending and improving the available data on faculae and network regions, critical for an assessment of the long-term changes in solar activity and irradiance. To allow reconstructions yet further back in time, e.g., from sunspot records, 
understanding of the detailed relationship between sunspot and facular evolution is pivotal. Various studies have addressed this issue \citep[e.g.][and references therein]{foukal_curious_1993,shapiro_variability_2014,yeo20a,chatzistergos_scrutinising_2022,nemec_faculae_2022}, generally confirming its non-linearity.
More recent results favour a power-law relationship between plage areas and the sunspot number as well as sunspot areas \citep{shapiro_variability_2014,yeo20a,chatzistergos_scrutinising_2022}.
The relationship has also been found to exhibit a weak dependence on the activity level  \citep{chatzistergos_scrutinising_2022}.
A direct consequence of the non-linearity of the relationship between plage areas and sunspot areas or numbers is the non-linearity of the relationship between the sunspot number or areas and the F10.7 radio flux \citep[see, e.g.,][]{yeo20a}. The latter has been often used by climate and atmosphere models and the question about the apparent divergence between sunspot records and F10.7 over the more recent period has been raised in the literature \citep[see][and references therein]{clette_is_2021}.

\section{The dimmest state of the Sun}
\label{sec:dimmeststate}

Current estimates of the change in the mean TSI since the Maunder minimum, listed in Table \ref{tab:irradiancereconstructions}, range from a rise by $5.5\ {\rm Wm^{-2}}$ to a drop by $-0.09\ {\rm Wm^{-2}}$. 
This uncertainty also spreads to the longer-term reconstructions, see the discussion in Sect.~\ref{sec:proxy_overview}.
These current estimates are based on empirical or semi-empirical models of solar irradiance variability.

A different approach was taken by \cite{yeo17b}.
This model employs three-dimensional (3D) solar model atmospheres based on realistic magnetohydrodynamic (MHD) simulations of the solar surface and atmosphere.
As a result, this model, which we will refer to as SATIRE-3D, does not require any calibration to measured solar irradiance variability. 
 This is currently the only model reported in the literature to achieve this. 
 This unique feature of the model allows an estimate of the lowest possible TSI level during periods of extremely low solar activity, such as during grand minima. A comparison with the current TSI levels then allows setting an upper limit on the possible secular TSI change.
Similarly to semi-empirical models, the \cite{yeo17b} model has two components. The first component is the solar disc coverage by faculae and network, classed together and termed collectively as faculae, and by sunspots. This is derived by identifying these features in full-disc longitudinal magnetograms and continuum intensity images from the Helioseismic and Magnetic Imager onboard the Solar Dynamics Observatory \citep[SDO/HMI;][]{scherrer12} space telescope. The solar disc outside of sunspots and faculae (including network) is classified as internetwork. 
The second component of the model is the intensity of sunspots, faculae and the internetwork. Sunspot intensity is calculated from plane-parallel model atmospheres \citep{unruh99}. Indeed, sunspots can be reasonably well represented by plane-parallel model atmospheres such that their incorporation into the model reproduces sunspot darkening without issue. However, in contrast to semi-empirical models, facular and internetwork components are represented by state-of-the-art 3D model atmospheres, generated with MHD simulations of facular and internetwork regions using the MURaM code \citep[MPS/University of Chicago Radiative MHD;][]{vogler04,vogler05,vogler07,rempel14,rempel20}. 
The latter allows a unique conversion of the magnetic field signal measured in a magnetogram into the brightness contrast of a given facular/network pixel \citep[see][for details]{yeo_solar_2017}.

\begin{figure*}[]
	\centering
		\begin{overpic}[width=0.9\linewidth,trim={0 0.cm 0cm 0.cm},clip]{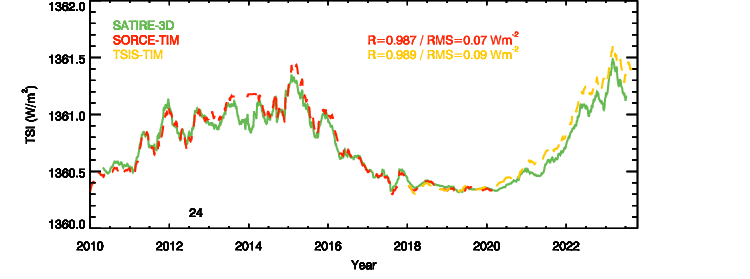}\put (13.00000,28.0000) {} \end{overpic}  
\begin{overpic}[width=0.9\linewidth,trim={0 0.0cm 0cm 0.cm},clip]{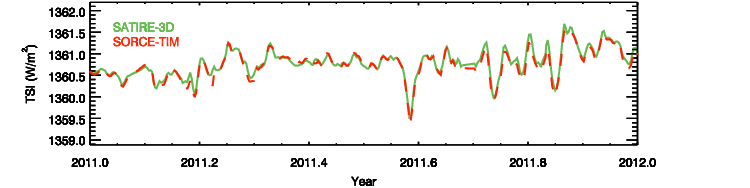}\put (14.00000,22.0000) { } \end{overpic}  
\begin{overpic}[width=0.9\linewidth,trim={0 0.0cm 0cm 0.cm},clip]{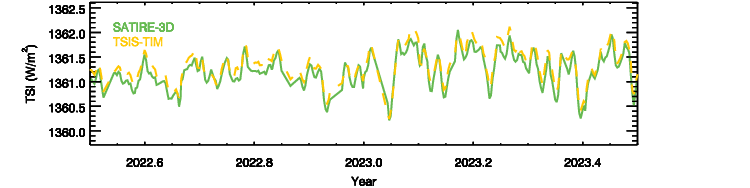}\put (14.00000,22.0000) { } \end{overpic}  
	  \caption{TSI reconstruction with the SATIRE-3D model (green) along with the SORCE/TIM (red) and TSIS/TIM (yellow) TSI measurements. The entire period covered by SATIRE-3D is shown in the top panel, while the other two panels show the series over a one-year interval starting in January 2011 and July 2022, respectively. 81-day running means are shown in the top panel and daily values in the other panels.  Also listed within the top panel is the linear correlation coefficient, $R$, and the RMS differences between SATIRE-3D and the series TSI measurements.}\label{fig:tsirecsatire3d}
\end{figure*}

For a given time, the appropriate intensity is then assigned to each point on the solar disc depending on whether it is inside a sunspot, faculae or the internetwork, as determined from HMI observations, and the summation of the result yields the corresponding TSI level. Taking daily HMI observations from 30 April 2010, when the instrument started regular operation, to 31 December 2019 as input into the model, \cite{yeo20b} reconstructed TSI variability over the intervening period. This TSI reconstruction reproduces more than 97\% of the observed TSI variability over the same period, as indicated by the coefficient of determination with the SORCE/TIM TSI record (see Fig.~\ref{fig:tsirecsatire3d} comparing SATIRE-3D model extended until 2023 to SORCE/TIM and TSIS/TIM observations; note that only SORCE/TIM data were included for comparison in the original paper, as TSIS/TIM was not operating yet).

Making use of the state-of-the-art SDO/HMI data for computing the daily TSI values, this model cannot, unfortunately, be extended back to compute irradiance variations over the entire period since the Maunder minimum.
Instead, by considering the Sun in its least possible active stage and comparing the TSI during such a state with the current levels, it sets a limit on the maximum possible TSI change between a grand minimum and now.
The solar magnetic field is maintained by the global dynamo \citep{charbonneau20} and the small-scale dynamo or SSD, which refers to the interaction between solar convection and magnetic flux that produces the ubiquitous internetwork magnetic field \citep{borrero17}.
Based on the dearth of sunspots during the Maunder minimum and recent studies suggesting that the SSD is untethered to the global dynamo \citep{lites11,lites14,buehler13,rempel14}, \cite{yeo20b} argued that at the Maunder minimum, the global dynamo was unusually weak, while the SSD and the internetwork magnetic field it supports were likely to be in a similar state as they are today. The most inactive state the Sun can possibly be in during grand solar minima is therefore one where the global dynamo is completely dormant and the internetwork magnetic field, looking as it does today, extends the entire solar surface. Accordingly, by assuming the entire solar surface was covered by only the internetwork magnetic field \citep[using the SSD model by][]{rempel14} during a grand minimum,
\cite{yeo20b} established the TSI level of the Sun in its least active state. 

The resulting TSI level was found to be $1358.3\pm0.7\ {\rm Wm^{-2}}$ or $2.0\pm0.7\ {\rm Wm^{-2}}$ below the 2019 level. 
For a more consistent comparison with the numbers for various reconstructions listed in Table~\ref{tab:irradiancereconstructions}, which refer to the difference between 1986 and 1700 (as some of the reconstructions do not cover the minimum in 2019), we convert this into the maximum $\sim2.2\pm0.7\ {\rm Wm^{-2}}$ increase from the Maunder minimum to 1986, using the estimate by \citet{montillet_data_2022} for the TSI change between 1986 and 2019 (see Table~\ref{tab:irradiancecomposites}).
Since the TSI level of the least active Sun corresponds to the lowest level possible during grand solar minima, the difference to the present level, represents the greatest rise in TSI since the Maunder minimum possible. 
We emphasise, however, that according to proxy data and other observations (e.g., cosmogenic isotope concentrations, sunspots seeings, auroral activity) the ``dimmest Sun'' condition was most likely not reached during the Maunder minimum  \citep[see][and references therein]{krivova_modelling_2021}, and thus the TSI change since then is expected to be lower than this limit. 
The constraint is shown as horizontal black line in Fig. \ref{fig:tsireconstructions}. Its uncertainty is marked by the blue shading, while the TSI levels clearly outside this constraint are shaded in red. This cap on the change in TSI since the Maunder minimum encloses all the current estimates except that reported by \cite{egorova18}. 
The reasons why the approach by \citealt{shapiro11} and \citealt{egorova18} lead to a significantly higher estimates are discussed in \citealt{yeo20b}.
This is a robust constraint on possible TSI variations, which is in agreement with, although less tight than, other such estimates from the literature \citep[e.g.][]{schrijver_minimal_2011,feulner_are_2011,lockwood_placing_2020,marchenko_relationship_2022}.

\section{Summary and conclusions}
\label{sec:summary}
Long and reliable records of solar activity and irradiance are of paramount importance for understanding the solar influence on Earth's climate.
The available direct TSI measurements only extend back to 1978 and come from a number of different instruments.
The data exhibit non-negligible inconsistencies, and there is some vagueness in how they are put together into a continuous record. 
As a result, the magnitude of the secular change (if any at all) in the solar irradiance over the period of measurements remains unclear.
The estimates of the activity minimum-to-minimum change from 1986 to 2009
in the available TSI composites range between -8.4 and 10 Wm$^{-2}$y$^{-1}$ (see Table~\ref{tab:irradiancecomposites} and Fig.~\ref{fig:tsicomposites}).
For the period since 1996 a marginally decreasing trend is reported, while most composites also evidence a marginally decreasing TSI over the entire satellite period. 
The latter is supported by observations of the solar surface magnetic field (Fig.~\ref{fig:magflux}) and \ca data (Fig.~\ref{fig:tsireconstructions_phsum}).

Models are used to extend the TSI record into the past and reconcile the inconsistencies of the direct measurements.
Models recover irradiance variations by accounting for the different contributions of solar surface magnetic regions, such as dark sunspots and bright faculae (plage) and network.
The main source of the uncertainty in the long-term irradiance reconstructions is the absence, until recently, of a reliable proxy of facular brightening.
Thus, the estimates of the TSI change between 1700 and 1986 by various models range between -0.09 and 5.5 Wm$^{-2}$, although
more recent reconstructions (roughly from the last decade) are in a better agreement with each other.
In particular,  \cite{yeo20b} used cutting-edge 3D MHD simulations of the solar atmosphere with the (only currently existing) physical model, SATIRE-3D, to estimate the minimum TSI level that can be reached when the Sun becomes extremely quiet, similarly to the state of the so-called grand minima (e.g., Maunder or Spörer minima, to name some of the most famous ones).
Such a state could be reached if the global dynamo in the Sun became dormant, and only the small-scale dynamo driven by the interaction of the convection with the solar magnetic field and generating the  ubiquitous internetwork magnetic field were acting. 
It is not possible to reconstruct the solar irradiance with this model over the period not covered by state-of-the-art high-resolution solar observations, but a comparison of the ``dimmest'' state of the Sun with current minima allows a constraint on the maximum possible change in TSI between such boundary conditions.
In this way, \cite{yeo20b} estimated the difference in the TSI between 2019 and the dimmest state of the Sun to be 2.0$\pm$0.7 Wm$^{-2}$. The \cite{yeo20b} constraint on the dimmest state of the Sun excludes ``extreme'' TSI reconstructions, such as those by \cite{hoyt_discussion_1993}, \citet{fligge_solar_2000-1}, or \citet{egorova18}. 
The reconstruction by \citet{penza_total_2022} lies just at the allowed limit.

Reconstructions to yet earlier times, up to the entire Holocene, have also been done. They rely on cosmogenic radioisotope data, which are only an indirect proxy of solar activity.
They thus require further assumptions and do not allow an independent estimate of the magnitude of the secular variability.
Their temporal resolution is also lower, typically decadal, while the quality of the data worsens further back in time. 
Although, using the recent unique $^{14}$C data from \citet{brehm_eleven-year_2021}, \citet{usoskin_solar_2021}
have been able to reconstruct the annual sunspot number over the period 971--1900, which has increased the number of known solar cycle from 36 covered by telescopic observations to 96.

Huge efforts have also been invested into improving and extending the available historical proxies of solar magnetic activity, which is an ongoing process and will keep providing improved records for solar irradiance studies.
This includes significant revisions and
updates of the sunspot number records, and a great progress in exploiting the historical full-disc photographs of the Sun in the \ca line, which offer unique information on faculae since the late 19th century.

In particular, significant efforts have been put over the last two decades into digitisation of the available historical archives and development of accurate techniques for their processing and analysis \citep{chatzistergos_full-disc_2022}.
By now more than 43 archives have been analysed \citep{chatzistergos_analysis_2020}, providing an excellent temporal coverage back to 1892.
The potential of historical \ca observations for irradiance reconstructions with a semi-empirical model was demonstrated by \cite{chatzistergos_reconstructing_2021-1} paving the way for utilising these data for a more accurate TSI reconstruction back to 1892 than currently possible.

Some extra work to resolve remaining issues with \ca data is, however, still needed on the way to this goal. One of the most critical challenges is to account for various inconsistencies in the \ca archives, to avoid spurious trends in the outcome \citep{chatzistergos_full-disc_2022}.

Such a significantly refined reconstruction over the last century would also help to calibrate the magnitude of the secular variability and thus
impose constraints on TSI reconstructions going yet further back in time.
Extending reconstructions further back in time additionally requires a better understanding of the link between the evolution of sunspots and faculae.
The latter can be done with modern high-resolution data, historical sunspot and facular observations, and simulations \citep[see, e.g.,][]{yeo20a,yeo_relationship_2021,krivova_modelling_2021, chatzistergos_scrutinising_2022,nemec_faculae_2022}.

As a final note, in this review we have focused on the long-term changes in the TSI and did not discuss the SSI variability. While irradiance variability is strongly wavelength-dependent, recovering long-term trends, which are in the focus of this review, faces essentially the same problems for both the TSI and SSI. We note though that SSI measurements are even more challenging than TSI measurements \citep[see, e.g,,][]{ermolli_recent_2013,yeo_uv_2015,deland_creation_2019,woods_overview_2021}.

	\begin{acknowledgements}
The authors thank Yasser Abduallah, Francesco Berrilli, Steven Dewitte, Tatiana Egorova, Valeri Fedorov, Victor Privalsky, Werner K. Schmutz for providing their data.
We thank the anonymous referees for the useful comments that helped improving this paper. 
The TSI reconstructions with SATIRE-3D, T, T2, M, and by \cite{chatzistergos_modelling_2020} as well as the CEA20 plage area composite, the CEA17 group sunspot number series, and the MEA20 sunspot area composite are available at \url{http://www2.mps.mpg.de/projects/sun-climate/data.html}.
ISNv2, HoSc98, and SvSc16 series are available at \url{https://www.sidc.be/silso/datafiles}.
\cite{bard_solar_2000,bard_comment_2007,lean_evolution_2000,steinhilber_total_2009,steinhilber_9400_2012} TSI series are available at \url{https://www.ncei.noaa.gov/pub/data/paleo/climate_forcing/solar_variability/}, while the \cite{svalgaard_no_2007} one at \url{https://web.archive.org/web/20230927071933/https://svalgaard.leif.org/research/Historical-TSI.xls}.
\cite{wang05}, \cite{vieira_evolution_2011}, and \cite{delaygue_antarctic_2011} TSI series are included in the supplementary material by \cite{schmidt_climate_2011}.
This work was supported by the German Federal Ministry of Education and Research (Project No. 01LG1909C) and by the
European Union's Horizon 2020 research and Innovation program under grant agreement No 824135 (SOLARNET).
TC thanks ISSI for supporting the International Team 417 "Recalibration of the Sunspot Number Series".
This research has made use of NASA's Astrophysics Data System.
	\end{acknowledgements}

\bibliographystyle{aa}
\bibliography{_biblio01,references}

\end{document}